\pdfoutput=1
\documentclass[12pt,a4paper]{article}

\usepackage{lineno}  % for line numbering during review
\usepackage{graphicx}  % to include figures (can also use other packages)
\usepackage{xspace} % To avoid problems with missing or double spaces after
\usepackage{color}
\usepackage{colortbl}
\usepackage{amsmath} % Adds a large collection of math symbols
\usepackage{ifthen} % for conditional statements
\usepackage{multirow}
\graphicspath{{./figs/}} % Make Latex search fig subdir for figures

\newboolean{pdflatex}
\setboolean{pdflatex}{true} % use this if using non-eps figures
\newboolean{articletitles}
\setboolean{articletitles}{true} % False removes titles in references

\newboolean{uprightparticles}
\setboolean{uprightparticles}{false} %Set to true to get roman particle symbols

\usepackage{amssymb}
\usepackage{amsfonts}
\usepackage{upgreek} % Adds in support for greek letters in roman typeset

\usepackage{hyperref}    % Hyperlinks in references

\usepackage[all]{hypcap} % Internal hyperlinks to floats.

\graphicspath{{figures/}}
\def\ux85 {\mbox{UX85}\xspace}

\def\babar  {\mbox{BaBar}\xspace}

\ifthenelse{\boolean{uprightparticles}}%
{

 \def\Ppsi        {\ensuremath{\uppsi}\xspace}

 \def\PDelta      {\ensuremath{\Delta}\xspace}                 
 \def\PXi      {\ensuremath{\Xi}\xspace}                 
 \def\PLambda      {\ensuremath{\Lambda}\xspace}                 
 \def\PSigma      {\ensuremath{\Sigma}\xspace}                 
 \def\POmega      {\ensuremath{\Omega}\xspace}                 
 \def\PUpsilon      {\ensuremath{\Upsilon}\xspace}                 
 
 \def\PB      {\ensuremath{\mathrm{B}}\xspace}                 
                  
 \def\PD      {\ensuremath{\mathrm{D}}\xspace}

 \def\PJ      {\ensuremath{\mathrm{J}}\xspace}                 
 \def\PK      {\ensuremath{\mathrm{K}}\xspace}

 \def\Pi      {\ensuremath{\mathrm{i}}\xspace}

 \def\Ps      {\ensuremath{\mathrm{s}}\xspace}

}
{

 \def\Ppsi        {\ensuremath{\psi}\xspace}                 
                  
 \mathchardef\PDelta="7101
 \mathchardef\PXi="7104
 \mathchardef\PLambda="7103
 \mathchardef\PSigma="7106
 \mathchardef\POmega="710A
 \mathchardef\PUpsilon="7107
                  
 \def\PB      {\ensuremath{B}\xspace}                 
                  
 \def\PD      {\ensuremath{D}\xspace}

 \def\PJ      {\ensuremath{J}\xspace}                 
 \def\PK      {\ensuremath{K}\xspace}

 \def\Pi      {\ensuremath{i}\xspace}

 \def\Ps      {\ensuremath{s}\xspace}

}

\def\squark    {\ensuremath{\Ps}\xspace}

\def\kaon  {\ensuremath{\PK}\xspace}
  \def\Kbar  {\kern 0.2em\overline{\kern -0.2em \PK}{}\xspace}

\def\Kz    {\ensuremath{\kaon^0}\xspace}
\def\Kzb   {\ensuremath{\Kbar^0}\xspace}
\def\KzKzb {\ensuremath{\Kz \kern -0.16em \Kzb}\xspace}
\def\Kp    {\ensuremath{\kaon^+}\xspace}
\def\Km    {\ensuremath{\kaon^-}\xspace}

\def\KpKm  {\ensuremath{\Kp \kern -0.16em \Km}\xspace}

  \def\Dbar    {\kern 0.2em\overline{\kern -0.2em \PD}{}\xspace}
\def\D       {\ensuremath{\PD}\xspace}

\def\Dz      {\ensuremath{\D^0}\xspace}
\def\Dzb     {\ensuremath{\Dbar^0}\xspace}
\def\DzDzb   {\ensuremath{\Dz {\kern -0.16em \Dzb}}\xspace}
\def\Dp      {\ensuremath{\D^+}\xspace}
\def\Dm      {\ensuremath{\D^-}\xspace}

\def\DpDm    {\ensuremath{\Dp {\kern -0.16em \Dm}}\xspace}

\def\B       {\ensuremath{\PB}\xspace}
  \def\Bbar    {\kern 0.18em\overline{\kern -0.18em \PB}{}\xspace}

\def\Bz      {\ensuremath{\B^0}\xspace}
\def\Bzb     {\ensuremath{\Bbar^0}\xspace}

\def\Bs      {\ensuremath{\B^0_\squark}\xspace}
\def\Bsb     {\ensuremath{\Bbar^0_\squark}\xspace}

\def\jpsi     {\ensuremath{{\PJ\mskip -3mu/\mskip -2mu\Ppsi\mskip 2mu}}\xspace}

  \def\Y#1S{\ensuremath{\PUpsilon{(#1S)}}\xspace}% no space before {...}!

\def\Lbar {\ensuremath{\kern 0.1em\overline{\kern -0.1em\PLambda}}\xspace}

\def\to                 {\ensuremath{\rightarrow}\xspace}

\def\CP                {\ensuremath{C\!P}\xspace}

\def\AT#1     {\ensuremath{A_{\mathrm{T}}^{#1}}\xspace}           % 2

\def\C#1      {\ensuremath{\mathcal{C}_{#1}}\xspace}                       % 9
\def\Cp#1     {\ensuremath{\mathcal{C}_{#1}^{'}}\xspace}                    % 7
\def\Ceff#1   {\ensuremath{\mathcal{C}_{#1}^{\mathrm{(eff)}}}\xspace}        % 9  
\def\Cpeff#1  {\ensuremath{\mathcal{C}_{#1}^{'\mathrm{(eff)}}}\xspace}       % 7
\def\Ope#1    {\ensuremath{\mathcal{O}_{#1}}\xspace}                       % 2
\def\Opep#1   {\ensuremath{\mathcal{O}_{#1}^{'}}\xspace}                    % 7

\newcommand{\ket}[1]{\ensuremath{|#1\rangle}}              % {b}
\newcommand{\tev}{\ensuremath{\mathrm{\,Te\kern -0.1em V}}\xspace}
\newcommand{\gev}{\ensuremath{\mathrm{\,Ge\kern -0.1em V}}\xspace}
\newcommand{\mev}{\ensuremath{\mathrm{\,Me\kern -0.1em V}}\xspace}
\newcommand{\kev}{\ensuremath{\mathrm{\,ke\kern -0.1em V}}\xspace}
\newcommand{\ev}{\ensuremath{\mathrm{\,e\kern -0.1em V}}\xspace}
\newcommand{\gevc}{\ensuremath{{\mathrm{\,Ge\kern -0.1em V\!/}c}}\xspace}
\newcommand{\mevc}{\ensuremath{{\mathrm{\,Me\kern -0.1em V\!/}c}}\xspace}
\newcommand{\gevcc}{\ensuremath{{\mathrm{\,Ge\kern -0.1em V\!/}c^2}}\xspace}
\newcommand{\gevgevcccc}{\ensuremath{{\mathrm{\,Ge\kern -0.1em V^2\!/}c^4}}\xspace}
\newcommand{\mevcc}{\ensuremath{{\mathrm{\,Me\kern -0.1em V\!/}c^2}}\xspace}

\def\gsim{{~\raise.15em\hbox{$>$}\kern-.85em
          \lower.35em\hbox{$\sim$}~}\xspace}
\def\lsim{{~\raise.15em\hbox{$<$}\kern-.85em
          \lower.35em\hbox{$\sim$}~}\xspace}

\def\tell1  {TELL1\xspace}
\def\ukl1   {UKL1\xspace}

\newcommand{\etal}{{\slshape et al.\/}\xspace}

\usepackage{cite}
\usepackage{mciteplus}

\usepackage{epsfig,accents}

\begin{document}
\renewcommand{\thefootnote}{\fnsymbol{footnote}}
\setcounter{footnote}{1}
%%%%%%%%%%%%%%%%%%%%%%%%%
%%%%% Title     %%%%%%%%%
%%%%%%%%%%%%%%%%%%%%%%%%%

% %%%%%%% CHOOSE --------
%\input{title-LHCb-ANA}
%\input{title-LHCb-CONF}
%\input{title-LHCb-PAPER}
% %%%%%%%%%%%%% ---------

% $Id: title-LHCb-ANA.tex 10674 2011-10-13 13:13:10Z uegede $
% ===============================================================================
% Purpose: LHCb-ANA Note title page template
% Author:
% Created on: 2010-10-05
% ===============================================================================

%%%%%%%%%%%%%%%%%%%%%%%%%
%%%%%  TITLE PAGE  %%%%%%
%%%%%%%%%%%%%%%%%%%%%%%%%
\begin{titlepage}
\pagenumbering{roman}
% Header ---------------------------------------------------
\belowpdfbookmark{Title page}{title}

\pagenumbering{roman}
%\vspace*{-1.5cm}
%\centerline{\large EUROPEAN ORGANIZATION FOR NUCLEAR RESEARCH (CERN)}
%\vspace*{1.5cm}
\hspace*{-5mm}\begin{tabular*}{16cm}{lc@{\extracolsep{\fill}}r}
%\vspace*{-12mm}\mbox{\!\!\!\includegraphics[width=.12\textwidth]{lhcb-logo}}& & \\
%& & %LHCb-ANA-2012-xxx \\  % ID
 & &May 28, 2013 \\ % Date - Can also hardwire e.g.: 23 March 2010
% & & SU-HEP-010 \\
 & & \\
\hline
\end{tabular*}

\vspace*{2.5cm}

% Title --------------------------------------------------
{\bf\boldmath\huge
\begin{center}
Use of $B\to\jpsi f_0$ decays to discern the $q\bar{q}$ or tetraquark nature of scalar mesons
\end{center}
}

\vspace*{2.0cm}
% Authors -------------------------------------------------
\begin{center}
Sheldon Stone and Liming Zhang 
\bigskip\\
{\it\footnotesize
Physics Department
Syracuse University, Syracuse, NY, USA 13244-1130\\

}
\end{center}

%\vspace{\fill}

% Abstract -----------------------------------------------
\begin{abstract}
  \noindent
We consider the relative decay rates of  \Bzb and \Bsb mesons into a \jpsi plus a light scalar meson either the $f_0(500)$ ($\sigma$) or the $f_0(980)$.  We show that it is possible to distinguish between the quark content of the scalars being quark-antiquark or tetraquark by measuring specific ratios of decay rates. Using current data we determine the ratio of form-factors in $\Bsb\to\jpsi f_0(980)$  with respect to  $\Bzb\to\jpsi f_0(500)$ decays to be $0.99^{+0.13}_{-0.04}$ at a four-momentum transfer squared equal to the mass of the \jpsi meson squared. In the case where these light mesons are considered to be quark-antiquark states, we give a determination of the mixing angle between strange and light quark states of less than 29$^{\circ}$ at 90\% confidence level. We also discuss the use of a similar ratio to investigate the structure of other isospin singlet states.

%Total 5 variables ($t,m_{hh},\angmu,\angpi,\chi$) are used.

\end{abstract}

\vspace*{2.5cm}
%\vspace{\fill}

%\vspace*{1.0cm}
%{\it Keywords:} LHC, \CP violation, Hadronic $B$ Decays, $\Bsb$ meson\\
%\hspace*{6mm}{\it PACS:} 13.25.Hw, 14.40.Nd, 11.30.Er\\
%\hspace*{6mm}Submitted to Physics Review D\\
\newpage
% Authors -------------------------------------------------

\end{titlepage}

\renewcommand{\thefootnote}{\arabic{footnote}}
\setcounter{footnote}{0}

\pagestyle{empty}  % no page number for the title

%%%%%%%%%%%%%%%%%%%%%%%%%%%%%%%%
%%%%%  EOD OF TITLE PAGE  %%%%%%
%%%%%%%%%%%%%%%%%%%%%%%%%%%%%%%%

%  empty page follows the title page ----

\setcounter{page}{2}
\mbox{~}

%\cleardoublepage

%%%%%%%%%%%%%%%%%%%%%%%%%%%%%%%%
%%%%%  Table of Content   %%%%%%
%%%%%%%%%%%%%%%%%%%%%%%%%%%%%%%%
%%%% Uncomment next 2 lines if desired
%\tableofcontents
%\cleardoublepage

%%%%%%%%%%%%%%%%%%%%%%%%%
%%%%% Main text %%%%%%%%%
%%%%%%%%%%%%%%%%%%%%%%%%%

\pagestyle{plain} % restore page numbers for the main text
\setcounter{page}{1}
\pagenumbering{arabic}

% %%%%%%% CHOOSE --------
%% ----------------------------------
%% Line numbering on the left margin
%% ----------------------------------
%% Uncomment during review phase.
%% Comment it out before a final submission.
%\linenumbers
%% --------------------------------
% %%%%%%%%%%%%% ---------

% You can include short sections directly in the main tex file.
% However, for larger papers it is desirable to split the text into
% several semiautonomous files, which can be revised independently.
% This is especially useful when developing a document in
% collaboration with several people, since then different parts can be
% edited independently.  This type of file organization is shown here.
%

%f1(1285) mixing angle
Scalar mesons in general, and the $f_0(980)$ in particular are not well understood. Their masses do not follow the expectation in the na\"ive quark model that the state containing two strange quarks is heavier than the state containing only one, in stark contrast to the vector mesons \cite{PDG}. This has led to theories that the light $J^{PC}$ equal to $0^{++}$ mesons may be combinations of di-quarks and anti-diquarks,  e.g. $[qq][\bar{q}\bar{q}]$, called ``tetraquarks"  \cite{Hooft:2008we,*Weinberg:2013cfa,*Fariborz:2009cq,*Mennessier:2010ij,*Jaffe:1976ig,*Achasov:2012kk,*Achasov:2010fh}.

Recently there have been several studies of the $f_0(980)$ in heavy meson decays, some in the charm system \cite{Ecklund:2009aa,*delAmoSanchez:2010yp}. Based on these data, the existence of the mode  $\Bsb\to \jpsi f_0(980)$ was predicted \cite{Stone:2008ak}, discovered by the LHCb collaboration \cite{Aaij:2011fx}, and confirmed  \cite{Li:2011pg,*Aaltonen:2011nk,*Abazov:2011hv}. The LHCb collaboration also found the decay $\Bzb\to \jpsi f_0(500)$, and set an upper limit on the decay $\Bzb\to \jpsi f_0(980)$ \cite{Aaij:2013zpt}. 
From now on the $f_0(500)$ meson will be designated as $\sigma$, and the  $f_0(980)$  meson will be designed as $f_0$.   The $\Bsb\to \jpsi f_0(980)$ channel has also been used to measure \CP violation \cite{Aaij:2013oba}, but
Fleischer \etal have claimed that if the $f_0$ is a tetraquark state the measurement could be influenced  by the presence of additional  suppressed decay mechanisms \cite{Fleischer:2011au}.  Thus, a resolution of the problem of these states' structure would be helpful in several ways.

When the $\sigma$ and $f_0$ are considered as $q\bar{q}$ states there is the possibility of their being mixtures of light and strange quarks that is characterized by
a 2$\times$2 rotation matrix with a single parameter, the angle $\phi$, so that their wave-functions are 
\begin{eqnarray}
  \label{eq:fmix}
 \ket{f_0}&=&\;\;\;\cos\phi\ket{s\bar{s}}+\sin\phi\ket{n\bar{n}}\nonumber\\
  \ket{\sigma}&=&-\sin\phi\ket{s\bar{s}}+\cos\phi\ket{n\bar{n}},\nonumber\\
  {\rm where~} \ket{n\bar{n}}&\equiv&\frac{1}{\sqrt{2}}\left(\ket{u\bar{u}}+\ket{d\bar{d}}\right).
\end{eqnarray}

While there have been several attempts to measure the  mixing angle $\phi$, the model dependent results give a wide range of values.  We describe here only a few examples.  $D^{\pm}$ and $D_s^{\pm}$ decays into $f_0(980)\pi^{\pm}$ and $f_0(980)K^{\pm}$ give values of $31^{\circ}\pm5^{\circ}$ or $42^{\circ}\pm 7^{\circ}$ \cite{Black:1998wt}.  $D_s^+\to\pi^+\pi^+\pi^-$ transitions give a range $35^{\circ}<|\phi|<55^{\circ}$ \cite{Anisovich:2003up}. 
In light meson radiative decays two solutions are found either $4^{\circ}\pm3^{\circ}$ or $136^{\circ}\pm 6^{\circ}$ \cite{Anisovich:2001zp}.
Resonance decays from both $\phi\to\gamma\pi^0\pi^0$ and $\jpsi\to\omega\pi\pi$ give a value of $\simeq 20^{\circ}$.
On the basis of SU(3), a value of $19^{\circ}\pm5^{\circ}$ is provided \cite{Oller:2003vf}. Finally, Ochs \cite{Ochs:2013gi}, averaging over several processes, finds $30^{\circ}\pm3^{\circ}$

When these states are viewed as $q\bar{q}q\bar{q}$ states the  wave functions becomes 
\begin{equation}
 \ket{f_0}=\frac{1}{\sqrt{2}}\left([su][\bar{s}\bar{u}]+[sd][\bar{s}\bar{d}]\right),~~  \ket{\sigma}=[ud][\bar{u}\bar{d}].
 \end{equation}
In this Letter we assume the tetraquark states are unmixed, for which there is some justification \cite{Hooft:2008we,Black:1998wt,Maiani:2004uc},  with a mixing angle estimate of $<5^{\circ}$ \cite{Fleischer:2011au}.

In general, the decay width for a $\overline{B}$ meson to decay into a \jpsi and light scalar state $f$ can be expressed as \cite{ElBennich:2008xy,Li:2012sw,Fleischer:2011au}
\begin{equation}
\Gamma\left(\overline{B}\to \jpsi f\right)=C|F^f_{B}(m^2_{\jpsi})|^2|V_{ci}|^2\Phi {\cal{Z}}^2, 
\end{equation}
where $C$ is  a constant,  $F^f_B$ is form-factor evaluated at the four-momentum transfer $q^2$ equal to the mass of the \jpsi squared, and  $V_{ci}$ is the relevant CKM element. 
The phase space factor $\Phi=(m_B E(x,y))^3$, where $x=m_{\jpsi}/m_B$, $y=(m_f/m_B)$ and $E(x,y)=\sqrt{\left[1-(x+y)^2\right]\left[1-(x-y)^2\right]}$.\footnote{The phase space is calculated taking into account the mass dependent line shapes.} ${\cal{Z}}$ represents the coupling amplitude that depends on the quark configuration after the $\overline{B}$ meson decay, and the quark content of the light meson in either the $q\bar{q}$ or tetraquark model. The values for ${\cal{Z}}$ are listed in Table~\ref{tab:Z}.

\begin{table}[h!t!p!]
\centering
\caption{Values of the coupling amplitude ${\cal{Z}}$.}
\vspace{0.2cm}
\begin{tabular}{lcc|cc}
\hline
& \multicolumn{2}{c|}{\Bsb} &  \multicolumn{2}{c}\Bzb \\%\hline
Model  &$f_0$ & $\sigma$  & $f_0$ & $\sigma$ \\
\hline
$q\bar{q}$ & $\cos\phi$ & $\sin\phi$ & $\sin\phi/\sqrt{2}$& $\cos\phi/\sqrt{2}$\\
tetraquark & $\sqrt{2}$ & 0 &$1/\sqrt{2}$ & 1\\
\hline    
\end{tabular}
\label{tab:Z}
\end{table}

\begin{figure}[htb]
\vskip -.4cm
\begin{center}
\includegraphics[width=6in]{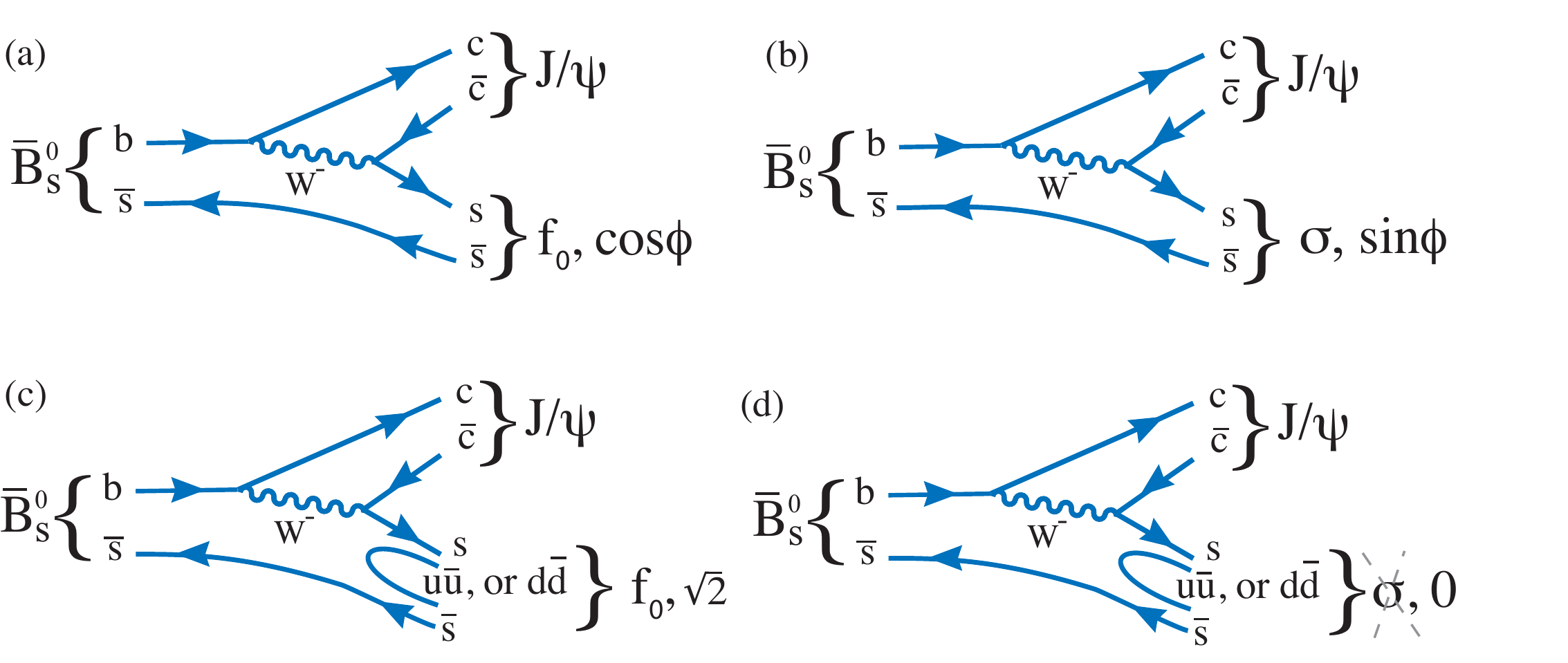}
\end{center}\label{jpsi-f0-sig-Bs}
\vskip -0.8cm
\caption{Decays of the \Bsb meson  to a \jpsi and  (a) $f_0$ in the $q\bar{q}$ model, (b) $\sigma$ in the $q\bar{q}$ model, (c) $f_0$ in the tetraquark model, and (d) $\sigma$ in the tetraquark model. The factor next to the scalar resonance name indicates the coupling amplitude ${\cal{Z}}$.}
\end{figure}

The diagrams for decays of \Bsb mesons into the $\sigma$ and $f_0$ are shown in Fig.~\ref{jpsi-f0-sig-Bs} for both $q\bar{q}$ and tetraquark models. The coupling amplitudes for the $f_0$, and $\sigma$ in the $q\bar{q}$ model are $\cos\phi$ and $\sin\phi$, respectively, while in the tetraquark model the coupling is $\sqrt{2}$ for the $f_0$ and $\sigma$ production is not allowed. Thus, a null test of the tetraquark model is evident: if the decay $\Bsb\to\jpsi \sigma$ is observed then the tetraquark model described here is ruled out. 

\begin{figure}[htb]
\vskip -.4cm
\begin{center}
\includegraphics[width=6in]{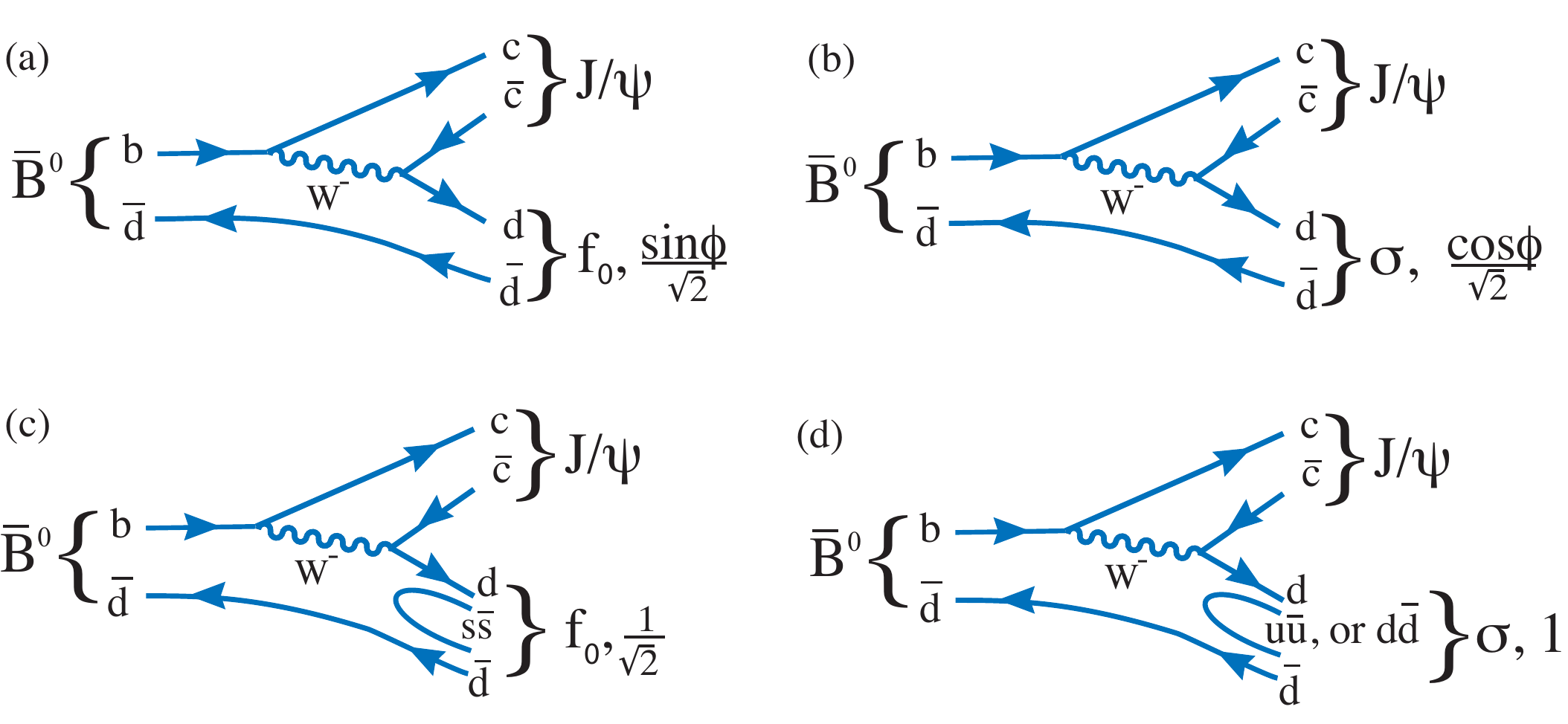}
\end{center}\label{jpsi-f0-sig-B0}
\vskip -0.8cm
\caption{Decays of the \Bzb  meson to a \jpsi and  (a) $f_0$ in the $q\bar{q}$ model, (b) $\sigma$ in the $q\bar{q}$ model, (c) $f_0$ in the tetraquark model, and (d) $\sigma$ in the tetraquark model. The factor next to the scalar resonance name indicates the coupling amplitude ${\cal{Z}}$.}
\end{figure}

The diagrams for decays of \Bzb mesons into the $\sigma$ and $f_0$ are shown in Fig.~\ref{jpsi-f0-sig-B0} for both $q\bar{q}$ or tetraquark models \cite{Wang:2009azc}.

There are measured branching fractions for some of these decays, that are summarized in Table~\ref{tab:data} .\footnote{In order to minimize systematic uncertainties we use only LHCb measurements even though other measurements of ${\cal{B}}(\Bsb\to\jpsi f_0)$ are available \cite{PDG}.}  
 The branching fractions into final states with an $f_0$  have been corrected by their decay rates into $\pi^+\pi^-$ using measurements from BES  \cite{Ablikim:2004cg,*Ablikim:2005kp} from which we obtain $\frac{{\cal{B}}\left(f_0(980)\to K^+K^-\right)}{{\cal{B}}\left(f_0(980)\to \pi^+\pi^-\right)}=0.25^{+0.17}_{-0.11}$\cite{Ecklund:2009aa}, and from \babar of $\frac{{\cal{B}}\left(f_0(980)\to K^+K^-\right)}{{\cal{B}}\left(f_0(980)\to \pi^+\pi^-\right)}=0.69\pm0.32$ \cite{Aubert:2006nu}.
 Averaging the two measurements gives
\begin{equation}
\frac{{\cal{B}}\left(f_0(980)\to K^+ K^-\right)}{{\cal{B}}\left(f_0(980)\to\pi^+\pi^-\right)}=0.35_{-0.14}^{+0.15}\,.
\end{equation}

To determine the $\pi^+\pi^-$ branching fraction it is assumed that the $\pi\pi$ and $KK$ decays are dominant, and that the ratios of $\pi^0\pi^0$ to $\pi^+\pi^-$, and $K^0\overline{K}^0$ to $K^+K^-$ are given by isospin conservation as 1/2 and 1, respectively, leading to \cite{Aaij:2013zpt}
\begin{equation}
{\cal{B}}\left(f_0(980)\to\pi^+\pi^-\right)
=\left(46\pm6\right)\%. 
\end{equation}
For $\sigma$ decay we use ${\cal{B}}\left(\sigma\to\pi^+\pi^-\right)
=\frac{2}{3}$, which again results from isospin conservation and the assumption that the only decays are into two pions. The uncertainties in these rates are not included in Table~\ref{tab:data}, but are introduced when comparisons between $\sigma$ and $f_0$ are made. 

\renewcommand{\arraystretch}{1.2}
\begin{table}[h!t!p!]
\centering
\caption{Experimental branching fractions from LHCb for $B\to \jpsi f$ meson final states. The uncertainties on ${\cal{B}}(f\to\pi^+\pi^-$) are not included.}
\vspace{0.2cm}
\begin{tabular}{lcc}
\hline
Final state& \Bsb \cite{LHCb:2012ae} & \Bzb \cite{Aaij:2013zpt} \\
\hline
$\sigma$ & $-$ & $9.60^{+3.79}_{-1.70} \times 10^{-6}$  \\
$f_0$ & $3.40^{+0.63}_{-0.16}\times 10^{-4}$  &  $<1.7\times 10^{-6}$ \\
\hline    
\end{tabular}
\label{tab:data}
\end{table}

In this Letter we present information obtainable from ratios of the $\Bsb$ and $\Bzb$ decay rates into $\sigma$ and $f_0$ mesons. Using the ratios allows cancellation of many of the experimental and theoretical uncertainties. The ratios we will consider are listed in Table~\ref{tab:rats}
for both $q\bar{q}$ and tetraquark models.
\begin{table}[h!t!p!]
\centering
\caption{Ratios of decay widths. The rate ratio  is multiplied by the value for  ${\cal{Z}}^2$ in either the $q\bar{q}$ model, or  the tetraquark model. The form-factors are notated as $F^i_j$, and the phase space factor $\Phi_j^i$, where $i$ indicates either $\sigma$ or $f_0$ and $j$ indicates either $\Bzb$ or \Bsb .}
\vspace{0.2cm}
\begin{tabular}{lcccc}
\hline
Label &Mode ratio  &Rate ratio& ${\cal{Z}}^2$ $q\bar{q}$ & ${\cal{Z}}^2$  tetraquark \\\hline 
$r^{0f_0}_{sf_0}$ &\multicolumn{2}{l}{\Large $\frac{\Gamma(\Bzb\to \jpsi f_{0})}{\Gamma(\Bsb\to \jpsi f_{0})}=\frac{|F^{f_0}_{\Bz}(m^2_{\jpsi})|^2}{|F^{f_0}_{\Bs}(m^2_{\jpsi})|^2}\frac{|V_{cd}|^2\Phi^{f_0}_{\Bz}}{|V_{cs}|^2\Phi^{f_0}_{\Bs}}$} & $\frac{1}{2}\tan^2\phi$   &$\frac{1}{4}$\\
$r^{0f_0}_{0\sigma}$ &\multicolumn{2}{l}{\Large $\frac{\Gamma(\Bzb\to \jpsi f_{0})}{\Gamma(\Bzb\to \jpsi \sigma)}=\frac{|F^{f_0}_{\Bz}(m^2_{\jpsi})|^2}{|F^{\sigma}_{\Bz}(m^2_{\jpsi})|^2}\frac{\Phi^{f_0}_{\Bz}}{\Phi^{\sigma}_{\Bz}}$} & $\tan^2\phi$   &$\frac{1}{2}$\\
$r_{sf_0}^{s\sigma}$ &\multicolumn{2}{l}{\Large $\frac{\Gamma(\Bsb\to \jpsi \sigma)}{\Gamma(\Bsb\to \jpsi f_{0})}
=\frac{|F^{\sigma}_{\Bs}(m^2_{\jpsi})|^2}{|F^{f_0}_{\Bs}(m^2_{\jpsi})|^2}\frac{\Phi^{\sigma}_{\Bs}}{\Phi^{f_0}_{\Bs}}$}
 & $\tan^2\phi$   &0\\
$r^{sf_0}_{0\sigma}$ &\multicolumn{2}{l}{\Large $\frac{\Gamma(\Bsb\to \jpsi f_{0})}{\Gamma(\Bzb\to \jpsi \sigma)}=\frac{|F^{f_0}_{\Bs}(m^2_{\jpsi})|^2}{|F^{\sigma}_{\Bz}(m^2_{\jpsi})|^2}\frac{|V_{cs}|^2\Phi^{f_0}_{\Bs}}{|V_{cd}|^2\Phi^{\sigma}_{\Bz}}$} & 2   &2 \\
\hline   
\end{tabular}
\label{tab:rats}
\end{table}

To calculate the width ratios from the branching fractions when both \Bzb and \Bsb initial states are present, we use values of the lifetimes of 1.530$\pm$0.007~ps and 1.622$\pm$0.0023~ps \cite{HFAG}, respectively. (Since the \Bsb modes are all  negative \CP eigenstates, we use the value provided for $\tau_{\rm long}$.)
Input on the form-factor ratios is needed to reach quantitative conclusions. For $r^{sf_0}_{0\sigma}$ both the $q\bar{q}$ and tetraquark models predict identical ratios, and this ratio  is independent of $\phi$. Using the data in Table~\ref{tab:data} we find
\begin{equation}
\label{eq:ffus}
\frac{|F^{f_0}_{\Bs}(m^2_{\jpsi})|}{|F^{\sigma}_{\Bz}(m^2_{\jpsi})|}=0.99^{+0.13}_{-0.04}.
\end{equation}

The ratio  $r_{sf_0}^{s\sigma}$ was suggested as a way of measuring  $\tan\phi$ by Li \etal \cite{Li:2012sw}. The form-factor ratio calculated by Li \etal is very close to unity, ${|F^{\sigma}_{\Bs}(m^2_{\jpsi})|^2}/{|F^{f_0}_{\Bs}(m^2_{\jpsi})|^2}$ =1.  Assuming that the similar form-factor ratio ${|F^{f_0}_{\Bz}(m^2_{\jpsi})}|/{|F^{\sigma}_{\Bz}(m^2_{\jpsi})|}$ is unity,  LHCb used their data to set an upper limit on $\phi<31^{\circ}$ at 90\% confidence level \cite{Aaij:2013zpt}.

Measurement of the branching fraction of  $\Bzb\to \jpsi f_{0}$ was suggested by Fleischer \etal  \cite{Fleischer:2011au} as a way of investigating the tetraquark structure of the $f_0$. In the $q\overline{q}$ model they use the form-factor ratio, ${|F^{f_0}_{\Bz}(m^2_{\jpsi})|}/{|F^{f_0}_{\Bs}(m^2_{\jpsi})|}$ that was computed by El-Bennich \etal \cite{ElBennich:2008xy} of 0.69 using dispersion relations.\footnote{In the covariant light front dynamics model El-Bennich \etal compute 0.58.}  They find results that are mixing angle dependent. In the tetraquark model they use a unit form-factor ratio,
and predict  ${\cal{B}}(\Bzb\to \jpsi f_{0},~f_0\to\pi^+\pi^-)\sim(1-3)\times 10^{-6}$. The measured upper limit from LHCb is  $1.1\times 10^{-6}$ at 90\% c.l., which is barely consistent.  It is also interesting that using the upper limit on the measured ratio $r^{0f_0}_{sf_0}$  and a unit form-factor ratio, we find an upper limit $\phi<29^{\circ}$ in the $q\bar{q}$ model, slightly more restrictive than the LHCb determined limit of $\phi<31^{\circ}$  using $r_{sf_0}^{s\sigma}$; this evaluation does not depend on any properties of the $\sigma$, nor on ${\cal{B}}(f_0\to\pi^+\pi^-)$. The ratio  $r^{0f_0}_{sf_0}$ was also suggested by Ochs \cite{Ochs:2013gi} as a way of investigating the properties of the $f_0(980)$ and the $f_0(1500)$; he also takes a unit form-factor ratio. 

A further elucidation of  the null prediction of $r_{sf_0}^{s\sigma}$ in the tetraquark model is in order. Besides the caveat that there could be a small amount of mixing, $<5^{\circ}$, between the $\sigma$ and $f_0$ tetraquark states, there also could be higher order diagrams that couple to the $\sigma$ in the \Bsb decay. In terms of the topological diagrams illustrated in ref.~\cite{Fleischer:2011au}, both the tree and leading penguin diagrams don't couple to the $\sigma$,  as well as three other higher order diagrams. On the other hand three diagrams involving penguin annihilation and $W$ exchange would couple to the $\sigma$. As these are expected to have a very small in rate compared to the tree diagram, we do not expect that they could induce a rate corresponding to a mixing angle of more than a few degrees. 

In conclusion, we discuss the importance of branching fraction ratios in ($\Bsb$ or $\Bzb$)$\to \jpsi$($\sigma$ or $f_0$) decays. These measurements can discern whether or not the $\sigma$ and $f_0$ are $q\bar{q}$ or tetraquarks. To aid in these tests we have determined the form-factor ratio $\frac{|F^{f_0}_{\Bs}(m^2_{\jpsi})|}{|F^{\sigma}_{\Bz}(m^2_{\jpsi})|}=0.99^{+0.13}_{-0.04}$, based on LHCb data. If the $\sigma$ is a tetraquark state, we do not expect to see it $\Bsb$ decays at a level of more than a percent of the $f_0(980)$ rate. For the $\sigma$ and $f_0$ being $q\overline{q}$ states we provide a limit on the mixing angle of $<29^{\circ}$ at 90\% confidence level. Furthermore, we note that these tests could be extended to other systems. For example, if an isospin equal zero meson, $f_{I=0}$ was found in both $\Bzb\to\jpsi f_{I=0}$ and $\Bsb\to\jpsi f_{I=0}$ decays its mixing angle with another meson could be determined using a ratio similar to $r^{0f_0}_{sf_0}$ (See also ref.~\cite{Ochs:2013gi}). It is interesting that the square of the coupling amplitude would be 1/4 in the tetraquark model, and in the $q\bar{q}$ model its mixing angle with some other, possibly unknown, meson could be determined.

We gratefully acknowledge support from  the National Science Foundation, and thank Jack Laiho, and Joe Schechter for useful discussions.
\newpage
\ifx\mcitethebibliography\mciteundefinedmacro
\PackageError{LHCb.bst}{mciteplus.sty has not been loaded}
{This bibstyle requires the use of the mciteplus package.}\fi
\providecommand{\href}[2]{#2}

%\bibliographystyle{LHCb}
%\bibliography{f0-mix}

\end{document}